# Long distance measurement using single soliton microcomb

Jindong Wang[1†], Zhizhou Lu[2,3†], Weiqiang Wang[2†*], Fumin Zhang[1*], Jiawei Chen[1], Yang Wang[2,3], Xianyu Zhao[1], Jihui Zheng[1], Sai T. Chu[4], Wei Zhao[2,3], Brent E. Little[2,3], Xinghua Qu[1], Wenfu Zhang[2,3*]

[1]State Key Lab of Precision Measuring Technology & Instruments, Tianjin University, Tianjin, China.

[2]State Key Laboratory of Transient Optics and Photonics, Xi'an Institute of Optics and Precision Mechanics, Chinese Academy of Sciences, Xi'an 710119, China.

[3]University of Chinese Academy of Sciences, Beijing 100049, China.

[4]Department of Physics and Materials Science, City University of Hong Kong, Hong Kong, China.

*Correspondence to: wwq@opt.ac.cn (W. W.); zhangfumin@tju.edu.cn (F. Z.); wfuzhang@opt.ac.cn (W. Z.).

†These authors contribute equal to this work.

**Abstract:** Dispersive interferometry (DPI) takes a major interest in optical frequency comb (OFC) based long distance laser-based light detection and ranging (LIDAR) for the merits of strong anti-interference ability and long coherent length. However, the mismatch between the repetition rate of OFC and the resolution of optical spectrum acquisition system induces a large dead-zone which is a major obstacle for practical applications. Here, a new DPI LIDAR on the strength of high-repetition-rate soliton microcomb is demonstrated, which reaches a minimum Allan deviation of 27 nm for an outdoor 1179 m ranging experiment. The proposed scheme approaches a compact, high-accuracy, and none-dead-zone long distance ranging system, opening up new opportunities for emerging applications of frontier scientific researches and advanced manufacturing.

## Introduction

High-accuracy long distance ranging plays a significant role in frontier sciences, such as multi-satellite flying formations based extraterrestrial planets searching, black holes imaging and gravitational waves detection, which heavily rely on the precision of a real-time satellites position detection system *(1-4)*. High-accuracy long distance ranging is also highly demanded in current advanced manufacturing technology for industrial processing and precision guidance *(5-7)*. During the last decades, LIDAR takes a major interest in the scientific community for the high angle, distance and velocity resolution, high anti-interference capability, as well as compact volume *(8)*. Recently, OFCs are used as revolutionary laser sources to improve the ranging accuracy, acquisition speed and extending distances of LIDAR *(9-11)*, and many ranging techniques are developed, such as time-of-flight method *(12, 13)*, synthetic wavelength interferometry *(14-16)*, DPI *(17-21)*, dual-comb method *(22-24)* as well as their combinations *(25)*. DPI, also known as frequency-domain interferometry or spectral interferometry, is well suited for long distance ranging for the merits of long coherent length and high tolerance against interference *(19, 20)*. The schematic of Michelson interferometer based DPI LIDAR is shown in Fig. 1A. The distance information is demodulated from the interference spectrum envelop. The maximum measurable distance $l_{max}$ depends on the resolution of optical spectrum acquisition system. Once $l_{max}$ is less than the distance measurement cycle $l_{pp}/2$ (inversely proportional to the repetition rate of the OFC), a measurement dead-zone from $l_{max}$ to $l_{pp} - l_{max}$ will be introduced, as shown in Fig. 1B *(26)*. However, limited by the low repetition rate of conventional mode-locked laser based OFCs, complex experimental techniques, such as scanning reference arm length, repetition rate tuning method *(27)*, and dual-comb or tri-comb schemes *(28)*, are employed to avoid the dead zone. But they are at the expense of the real-time performance, as well as the stability and integration capacity, which are great challenges for practical DPI ranging systems.

Actually, for a given optical spectrum acquisition system, a simple and effective scheme is using a high-repetition-rate OFC to shorten the distance measurement cycle. When the repetition rate of OFC is higher than $2n_g\delta f$ ($n_g$ is the air group refractive index, $\delta f$ is the resolution of the optical spectrum acquisition system), the dead-zone will be eliminated as shown in Fig. 1B. The emergence of microresonator based soliton microcombs (SMCs), realizing double balances between nonlinearity and dispersion as well as parametric gain and cavity loss, advances a novel type of integrated broadband coherent optical sources *(27-32)*. Benefitted from the miniature structure of microresontors, SMCs are featured by the ultrahigh repetition rate, which have exhibited unprecedented prospects in the areas of classical and quantum optical communication systems *(33, 34)*, dual-comb spectroscopy *(35)*, chip-scale optical frequency synthesizers *(36)*, and high accuracy ranging *(37, 38)*, etc. Benefitting from the ultrahigh repetition rate, SMC is an ideal laser source to eliminate the dead-zone of DPI based long distance ranging and have great advantage for compact integration.

In this work, a SMC-based-DPI ranging system is developed for none-dead-zone long distance measurement. Meanwhile, the periodic ambiguity is eliminated using a parallel dual-frequency phase-modulated laser rangefinder (PLR). The feasibility of the proposed ranging system is demonstrated in two scenarios. In the first scenario, an 80 m distance is measured in a well-maintained environment with a minimum Allan deviation of 5.39 nm. For the second scenario, the ranging system is built in an outdoor environment for 1179 m distance measurement, and a minimum Allan deviation of 27 nm is achieved while the effect of the air refractive index fluctuation is neglected. The proposed ranging system could be extended from millimeter to hundreds of kilometers distance detection, and has the potential of miniature integration, aiming to the applications of robotics, autonomous vehicles, formation flying of satellites, advanced manufacturing and frontier scientific researches.

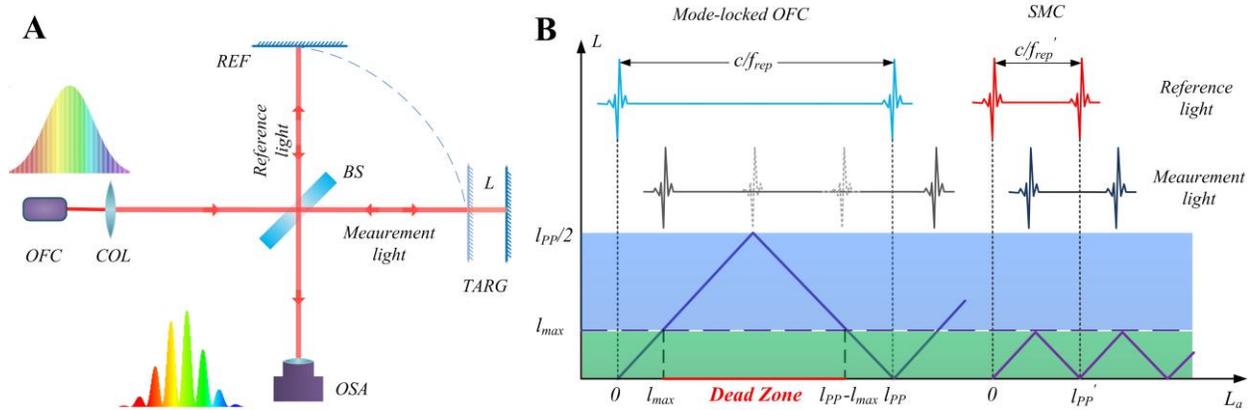

**Fig. 1 Principle of DPI ranging system.** **(A)** Schematic of Michelson interferometer based DPI ranging system. **(B)** The dead zone is introduced for the mismatch between the repetition rate of OFC and the resolution of optical spectrum acquisition system. Using a SMC with repetition rate

higher than $2n_g\delta f$, the dead zone will be eliminated. COL: Collimating mirror; BS: Beam splitter; REF: Reference mirror; OSA: Optical spectrum analyzer; TARG: Target mirror. $L_a$: Actual distance to be measured; $L$: Measured distance of a DPI system.

**Experimental setup**

The core SMC is generated in a high-index doped silica glass micro-ring resonator (MRR) with a quality factor of 1.7 million *(40, 41)*. Two key techniques are adopted to ensure SMC generate and survive in the unguaranteed experimental environment. First, the MRR is butterfly packaged with a compact thermal-electric cooler (TEC) (Fig. 2B). The TEC is used to tune the resonances of the MRR for SMC generation and isolate the temperature fluctuation of external environment to maintain SMC. Second, an auxiliary-laser-assisted intracavity thermal-balanced scheme is adopted to access single SMC in a deterministic fashion *(42)*. The experimental setup is shown in Fig. 2A. In our experiments, the wavelength of pump laser is fixed at 1560.2 nm with a linewidth of 100 Hz, and the pump and auxiliary lasers have a similar on-chip power of ~400 mW. Figure 2C shows a typical optical spectrum of single SMC which exhibits a standard squared hyperbolic secant (*sech*$^2$) envelope. The mode-crossing is almost avoided over the SMC bandwidth, which benefits from the stringent spatial-mode control of the device. The repetition rate of single SMC is 48.97 GHz, corresponding to a pulse spacing of 20.4 ps. The SMC provides an ideal broadband coherent laser source for long distance ranging.

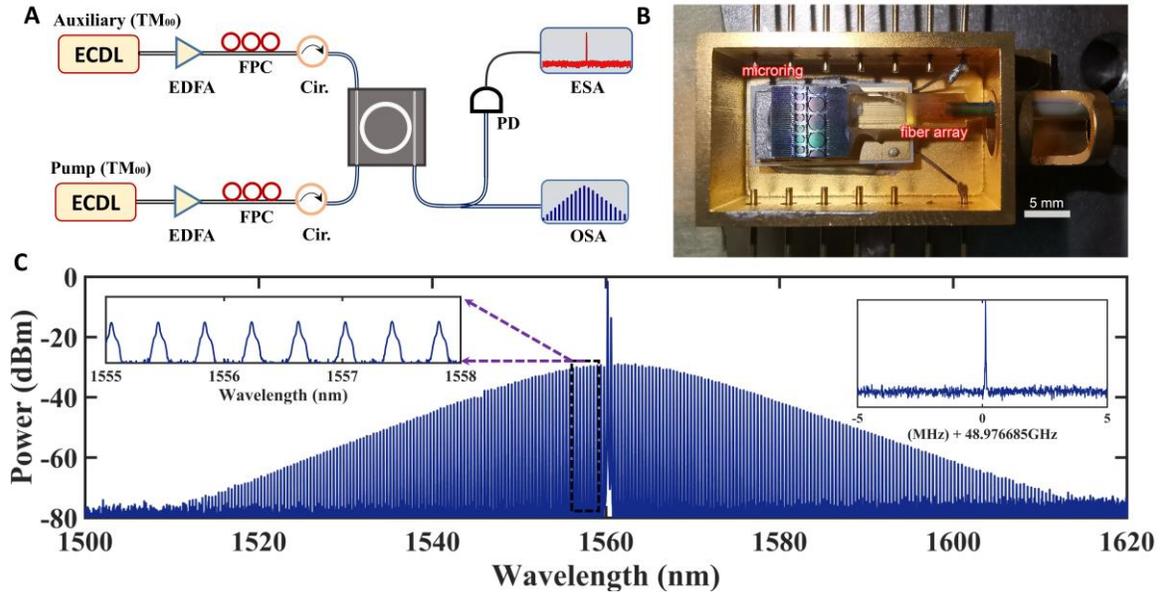

**Fig. 2. Single SMC generation in a high-index doped silica glass MRR. (A)** Experimental setup for SMC generation. An auxiliary-laser-assisted intracavity thermal-balanced scheme is adopted for single SMC generation. **(B)** Image of the butterfly packaged MRR. **(C)** A typical optical spectrum of the single SMC. Left inset: enlarged comb lines. Right inset: Electrical spectrum of single SMC. ECDL: External cavity diode laser; EDFA: Erbium-doped optical fiber amplifier; FPC: Fiber polarization controller; Cir: Circulator; PD: Photodetector; ESA: Electrical spectrum analyzer; OSA: Optical spectrum analyzer.

Figure 3 shows the experimental setup of the ranging system. For the DPI channel, a SMC is split into signal- and reference-comb using an 80:20 optical coupler (OC), and an optical attenuator (OA) is used to compensate the transmission loss of the signal comb to improve the visibility of interference fringe. The interference spectrum of the reflected signal comb and reference comb is spatially separated using two concave mirrors (CMs) and a reflection grating before lighting to a 256-pixels InGaAs linear array image sensor (LAIS) for interference fringe detection (Fig. 3A). The detection system is well designed to ensure that each LAIS pixel is illuminated by a single SMC mode (256 valid optical modes). The exposure time of the LAIS can be flexibly set between 26 to 260 μs by adjusting the integration time, which corresponds to a maximum measurement frequency of ~35 kHz considering the additional start-up time of 2.5 μs. An equivalent average process is realized during the spectrum acquisition which is helpful to alleviate the impact of white noise coupled from the pump source, EDFA and external environment. One typical interference spectrum is depicted in Fig. 3D. The distance information could be extracted from the spectrum data using an improved Fourier transform peak-to-peak method *(26)*. The Fourier transform spectrum (FTS) represents the delay time between the signal pulse and reference pulse as shown in Fig. 3F. In order to obtain the exact frequency peak of the FTS, a three-point fitting method is employed (Fig. 3G). Because of the ultrahigh repetition rate of the SMC, the ambiguity resolved distance is ~3.063 mm. In order to expand the measurable distance, a dual-frequency PLR is employed to estimate the distance, which is parallel with the SMC-based-DPI ranging system using a wavelength division multiplexer (WDM) technique. For the PLR channel, a phase modulated laser diode (LD) is used as light source, and the phase difference between the original modulation signal and the reflected laser signal is regained using a fast Fourier transform (FFT) digital phase discrimination technique. Through switching the LD modulation frequencies between 100 MHz and 100 kHz, the PLR realizes a measurement accuracy of ~1 mm over a measurable distance of 1500 m *(43)*. In our scheme, the distance under test is calculated by $L=N\times L_{am}+L_{frac}$, where $L_{am}$ and $L_{frac}$ are the ambiguity resolved distance and measured precise distance of the SMC-based-DPI ranging system, and $N$ is the integer when the PLR measured distance divided by the ambiguity resolved distance. The absolute distance can be real-time synthesized once the distance information from DPI and PLR channels is obtained simultaneously. To verify the capacity of our ranging system, we demonstrate the experimental implementation of an 80 m ranging in a well-maintained environment and a 1179 m ranging in an outdoor environment, respectively.

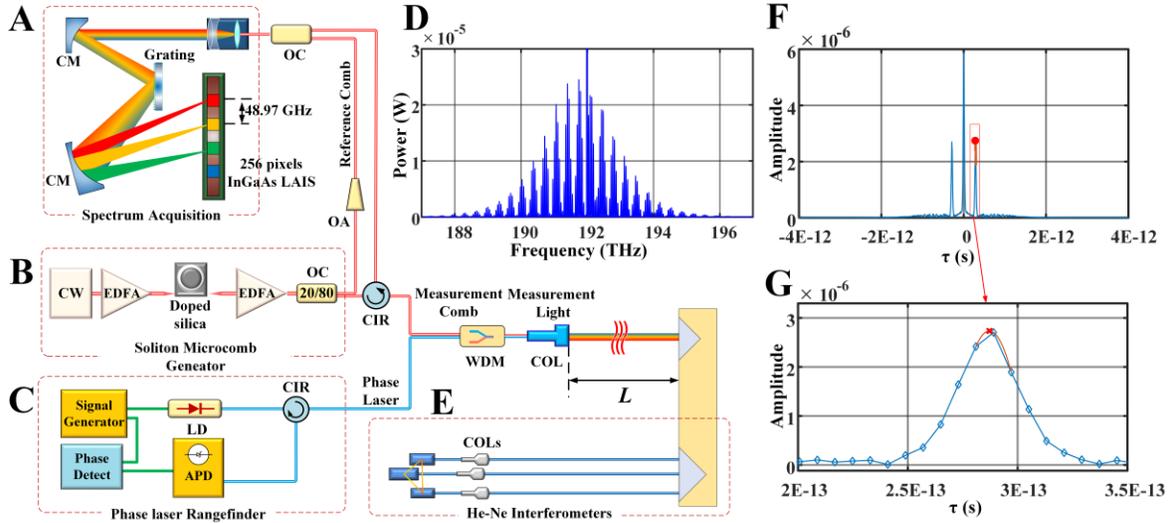

**Fig. 3. Schematic of the SMC-based-DPI ranging system. A dual-frequency PLR is employed for rough distance evaluation.** (**A**) Interference spectrum acquisition system. Two CMs and a reflection grating are designed to separate the interference spectrum spatially, which ensures each pixel of the LAIS is exposed by a single SMC mode. (**B**) SMC generator. A single SMC is split to signal and reference comb using a 20:80 OC. (**C**) Schematic of the PLR for distance evaluation. A phase modulated LD is used as optical source and the phase variation is demodulated by a phase detector. (**D**) Typical interference spectrum of the SMC-based-DPI ranging system. (**E**) An interferometer group, including three He-Ne laser interferometers to eliminate Abbe error, is used to provide a high-precision absolute distance reference. (**F**) FFT spectrum of the interference spectrum envelope. (**G**) Zoom-in vision of the FFT spectrum. A three-point fitting method (red line) is employed to extract the exact delay time. CM: Concave mirror; LAIS: Linear array image sensor; OC: Optical coupler; OA: Optical attenuator; CW: Continuous wave laser; EDFA: Erbium-doped optical fiber amplifier; CIR: Circulator; LD: laser diode; APD: Avalanche photodiode; COL: Collimating mirror; WDM: Wavelength division multiplexer.

**Ranging in a well-maintained environment**

The proposed SMC-based-DPI ranging scheme is firstly implemented in a well-maintained environment. The target is mounted on an 80 m precision granite guide rail system which locates in a 10 m deep basement at the National Institute of Metrology (NIM), China. The environmental parameters, including temperature, humidity, acoustical vibration and atmosphere pressure, are well monitored and controlled using the mounted sensor arrays and control systems, which results in minor environmental turbulence. The air refractive index is calibrated in real-time using the monitored environmental data based on the modified Edllen's formula *(43)*.

To obtain the absolute precision of the proposed ranging system, a referenced absolute distance is simultaneously measured using an interferometer group (IMG). The IMG consists of three parallelly installed He-Ne interferometers for Abbe error elimination (Fig. 3E). The absolute precision of the IMG is 0.07 μm + 0.7×10$^{-7}$ L in the controlled environment *(43)*. A reference point (zero distance) is calibrated by the proposed scheme and IMG ranging system synchronously. Then, the target mirror is moved from 0 to 3 mm with a step size of 0.05 mm. The measured results are shown in Fig. 4A (top), while the residuals between the measured and reference values are presented in Fig. 4A (bottom) with standard deviations error bars. The residuals are within 100 nm, which indicates the two ranging systems have a similar accuracy.

The accuracy of IMG ranging system is linearly decreasing as the distance increases. It is no longer suited as a reference for long distance ranging. Therefore, a standard deviation analysis is used for measurement accuracy assessment in our following experiments. For long distance ranging experiments, the target is placed around 1 m and 80 m, respectively, and the interference spectra are continuously acquired for 6 seconds. Based on a fast peak fitting algorithm and a timely distance demodulation technique *(44)*, the distance information can be timely recovered in a medium-performance FPGA. The measurement results of 1 m and 80 m ranging experiments are shown in Figs. 4B and 4C, and the standard deviations are 2.99 nm and 9.95 nm, respectively. The minimum Allan deviations of the two ranging experiments are 2.88 nm at average time of 2.74 s and 5.39 nm at average time of 2.055 s, respectively (Fig. 4D). It is noted that there are apparent distance drifts for 80 m distance measurements (Fig. 4C) which is mainly caused by environmental fluctuation. We believe that the ranging error could be further improved once the environmental fluctuation is well compensated.

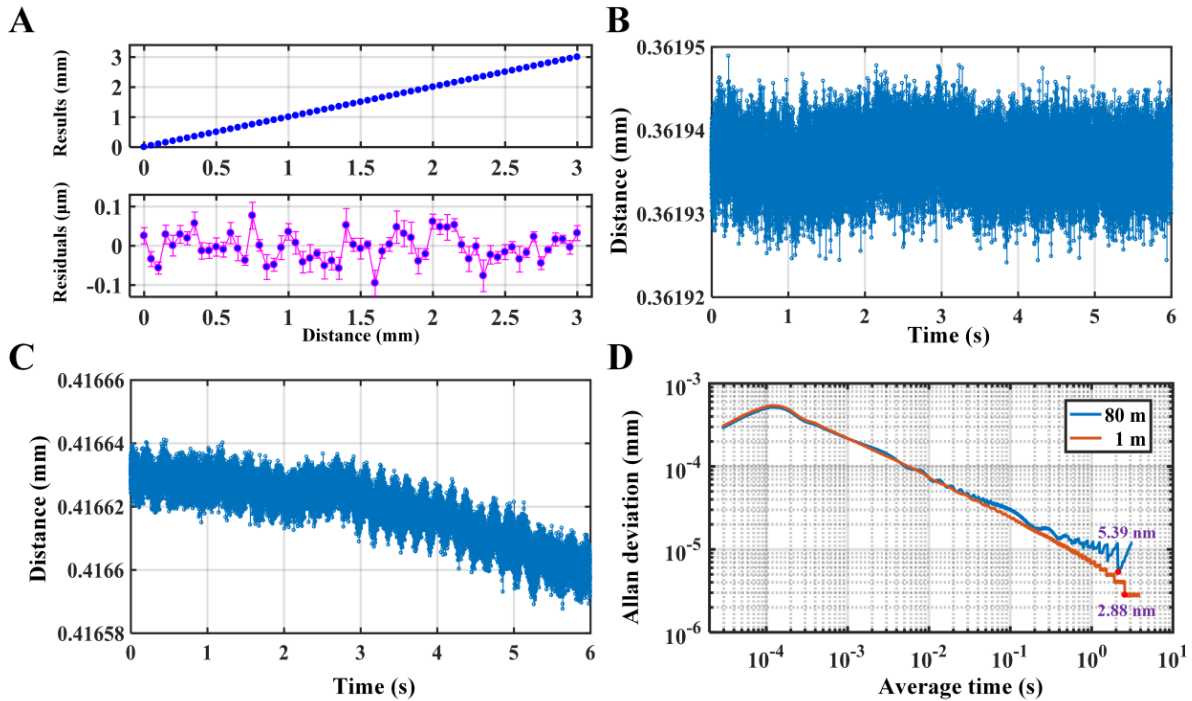

**Fig. 4. Experimental results of absolute distance ranging in a well-maintained environment.** **(A)** Measurement results when the target moves from 0 to 3 mm with a step size of 0.05 mm. Top: measured results versus set distance. Bottom: residuals between the DPI and IMG ranging results, and the standard deviations are labelled with error bars (magenta). **(B)** Measurement

results of 1 m ranging experiment. **(C)** Measurement results of 80 m ranging experiment. **(D)** Allan deviations versus average time of 1 m and 80 m ranging experiments.

**Long distance ranging in an outdoor environment**

The environmental applicability of the proposed ranging system is validated by building the system in a 1200 m outdoor standard baseline located at NIM (Fig. 5B). Environmental parameters along the baseline, including temperature, air pressure and humidity, are also monitored to calculate the air group refractive index *(43)*. To overcome the optical energy loss caused by beam expansion, an optical transmitter-receiver system is constructed, which consists of a Cassegrain telescope and an automatic space-to-fiber coupling system (Fig. 5A). And an eyepiece is used to collimate the light path and aim at the target. The automatic space-to-fiber system optimizes the coupling efficiency through a piezoelectric scanning method. The transmitting-receiving efficiency is about 1%, which is enough for our ranging experiments.

A cube-corner prism is used as the target and placed at about 1179 m away from the optical transmitter-receiver. In order to test the resolving ability of the proposed ranging system, the cube-corner prism is installed on a piezoelectric vibrator to provide high-frequency vibration signals, as shown in Fig. 5A. According to the test result of PLR, *N* is determined as 385,072. Considering the air effect group refractive index which is calculated to be 1.000,267,232 using the modified Edllen's formula based on the area weighted averaging parameters along the baseline, the stationary distance of the target is calibrated to 1,179,305.255,157,9 mm. The stationary absolute distance is continuously recorded using the proposed ranging system for 5 seconds, as shown in Fig. 5C. There is an obvious distance drift of ~100 μm, which is mainly induced by the low-frequency noise, arising from the temperature drift, environmental acoustic interference as well as air turbulence. The low-frequency noise can be clearly observed in the FTS (Fig. 5F) which is inevitable in an outdoor environment. According to the modified Edllen's formula, temperature is the main factor to the change of air refractive index (e.g., one degree temperature fluctuation will result in air refractive index changing by $10^{-7}$, corresponding to ~1 mm optical length drift for the distance of 1179 m). To alleviate the effect of low-frequency noise, a digital high-pass (HP) filter is employed. Figure 4E shows the Allan deviation of the stationary results before and after the HP filtering. Without the HP filter, the minimum Allan deviation is 5.6 μm at an average time of 0.2 ms, and the Allan deviation increases along with the average time increasing. Using a HP filter, the Allan deviation continuously decreases with increased average time and a minimum deviation of 27 nm is obtained at an average time of 1.8 s. 27 nm is regarded as the ranging capability of our outdoor ranging system, which is limited by the air dispersion which broadens the soliton pulse and leads to the spectral chirp. Such system would show better performance in a non-dispersive environment like outer-space where the air refractive index fluctuation and air dispersion are eliminated. Further, the capacity of high-frequency vibrations measurement is verified through imposing a modulation signal on the piezoelectric vibrator, e. g. two frequencies of 1 kHz and 5 kHz. The corresponding real-time ranging results are shown by the red and yellow lines in Fig. 5C, respectively. The zoom-in waveforms (Fig. 5D) represent vibrations of the target and the vibrational frequencies are clearly indicated by the Fourier transform spectra (Fig. 5F).

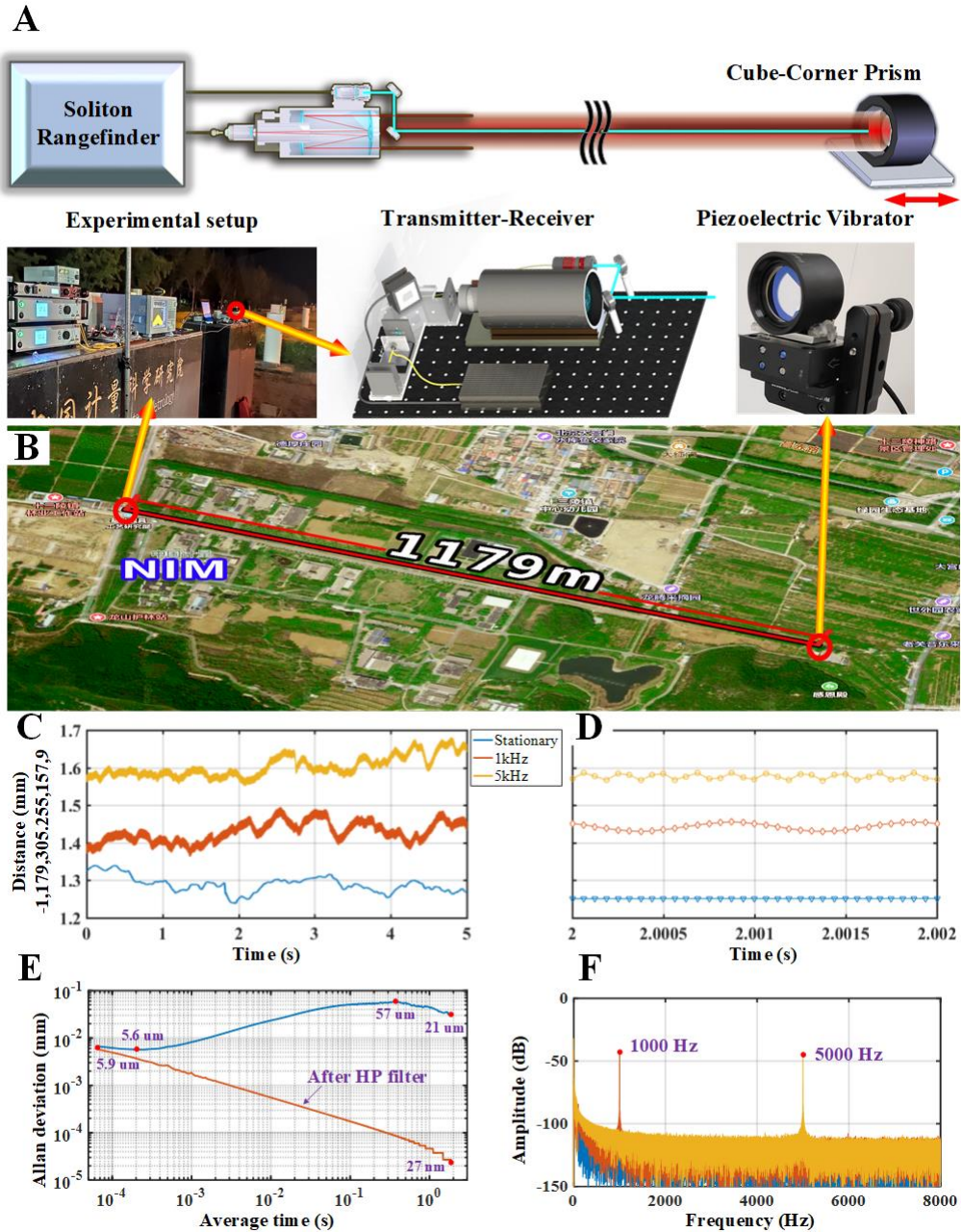

**Fig. 5. Experimental setup and results of long distance ranging in an outdoor environment.**
(**A**) Layout of the long distance ranging experiments. The emitted light is coaxial to the reflected light with the help of two reflectors. The lower left is a photography of experimental setup. A cube-corner prism is used as the target mirror and installed on a piezoelectric vibrator (lower right). (**B**) Satellite image of the outdoor baseline of NIM. The distance to be measured is ~1179 m. (**C**) Ranging results under three conditions: stationary (blue), 1 kHz vibration (red) and 5 kHz vibration (yellow). (**D**) Zoom-in of the ranging results. (**E**) Allan deviations of the stationary ranging results versus averaging time before and after a high-pass filter. (**F**) Frequency spectra under three conditions.

In conclusion, a real-time long distance ranging system is experimentally demonstrated using a high-repetition-rate single SMC as light source, the dead-zone is completely avoided and the distance detection accuracy can reach nanometer scale even in a practical outdoor environment for a distance over 1000 m. For the proposed scheme, the maximum ranging distance is determined by the available laser power as well as the coherent length of the laser source. The SMC is pumped by a 100 Hz linewidth laser which ensures the coherent length of the SMC over one thousand kilometers *(45)*. Therefore, such ranging system has the potential to extend the measurable distance to hundreds of kilometers for multi-satellite flying formations as long as the long-distance light transmitting-receiving antenna is available. Additionally, the ranging speed of our scheme is subject to the spectrum acquisition rate. Actually, each soliton pulse can act as an effective distance probe, thus the maximum achievable measuring speed can approach to the repetition rate of SMC (i.e., 48.97 GHz) once an ultra-high speed photodetector array is available. These excellent features, e.g., large dynamic range, ultra-high precision, extreme measuring speed and compact integration potential, ensure the SMC-based ranging system have extensive application prospects in both scientific and industrial areas.

**Acknowledgments:** This work is supported by the National Natural Science Foundation of China (Grant Nos. 51675380, 61675231, 51775379, 61635013, 61705257, 61805277), the National key research and development plan of China (Grant Nos. 2018YFF0212702, 2018YFB2003501), the Strategic Priority Research Program of the Chinese Academy of Sciences (Grant No. XDB24030600), the Youth Innovation Promotion Association, Chinese Academy of Sciences (Grant No. 2016353), and Key Projects Supported by Science and Technology of Tianjin, China (Grant No. 18YFZCGX00920). F. Z., W. W., W. Z., Wei. Z. and X. Q. organized the project. J. W., Z. Lu., J. C., Y. W. and X. Z. conducted the various experiments, J. C., J. W. and J. Z. analyzed the data. W. Z., W. W., S. C. and B. L. designed and fabricated the devices. W. W. packaged the samples. All authors discussed results and commented on the manuscript. J. W., Z. L., W. W. and W. Z. wrote the manuscript. W. Z., F. Z. supervised the project. We thank the support from prof. Mingzhao He of the National Institute of Metrology (NIM), China.